# What Have Google's Random Quantum Circuit Simulation Experiments Demonstrated about Quantum Supremacy?

Jack K. Horner, Lawrence KS 66046 USA
email: jhorner@cybermesa.com

John F. Symons, Department of Philosophy, University of Kansas,
Wescoe Hall, 1445 Jayhawk Boulevard, Lawrence KS  66045-7590 USA
email: johnfsymons@gmail.com

**Abstract**

*Quantum computing is of high interest because it promises to perform at least some kinds of computations much faster than classical computers.  Arute et al. 2019 (informally, "the Google Quantum Team") report the results of experiments that purport to demonstrate "quantum supremacy" – the claim that the performance of some quantum computers is better than that of classical computers on some problems.  Do these results close the debate over quantum supremacy?   We argue that they do not.  In the following, we provide an overview of the Google Quantum Team's experiments, then identify some open questions in the quest to demonstrate quantum supremacy.*

## 1.0   What is quantum supremacy?

Quantum computing is of high interest because it promises to perform at least some kinds of computations much faster than classical computers.  Quantum computers can execute some tasks "faster" than classical computers[1] only if those tasks can be executed "concurrently".[2]  For example, the search for prime numbers can be executed concurrently (see, for example, Bokhari 1984).  In contrast, solving some computational fluid dynamics systems requires time-ordering of computational steps  (see, for example, Kuzmin and Hämäläinen 2014) in the sense that there is no known computational method that would allow us to avoid that ordering.

---

[1] For the purpose of this paper, a "classical" computer is a computer to which a finite Turing Machine (Boolos, Burgess, and Jeffrey 2007, 44) is homomorphic.

[2] Informally speaking, a calculation-schema can be executed concurrently if more than one instance of the schema can be executed "at the same time" on a computing system. (For a more rigorous definition, see Hennessey and Patterson 2007, esp. Chaps. 2-4).

For the purposes of this paper, we need not comprehensively characterize classical or quantum computing; we assume, however, at least the following difference between quantum and classical computers (Nielsen and Chuang 2010, 13):

> (CQD) A *quantum computer* performs a computation in a way that is describable only by implying superposition of states as defined by the quantum theory (Messiah 1958; Bohm 1951). A *classical computer* in contrast performs a computation that is describable in a way that does not imply superposition of states.

Informally, *quantum supremacy* is the claim (see, for example, Harrow and Montanaro 2017; Boixo et al. 2018) that

> (IQS) A quantum computer can perform some set of computational tasks faster than a classical computer can.

There are many ways of interpreting IQS, depending on how we characterize "task", "faster", and "physical computation". The declared objective of Arute et al. 2019a (p. 505) is to demonstrate what we call *Google quantum supremacy*:

> (GQS) At least one computational task can be executed exponentially faster on a quantum computer than on a classical computer.

## 2.0 Overview of Arute et al. 2019

The computational task ("benchmark") used by the Google Quantum Team to demonstrate GQS is the sampling of the output of simulated pseudo-random quantum circuits. These simulated circuits are designed to entangle a set of quantum bits (qubits) by repeated application of single-qubit and two-qubit logical operations. The output of these circuits can be represented as a set of bitstrings (e.g., {000101, 11001, …}). Because of the entanglement induced by the operations on the qubits, some bitstrings are more much likely to occur than others (Arute et al. 2019a, 506).

Sampling the output of simulated random quantum circuits is a good benchmark for evaluating GQS because random circuits possess no structure other than the total number of qubits and the operations performed on individual, and pairs, of qubits. This approach helps to maximize the generality of the results, in the best case allowing the benchmark to confirm that the problem is computationally hard (e.g., cannot be solved in non-polynomial time).

The experiments of Arute et al. 2019 use a 53-qubit quantum computer – which they call the "Sycamore" processor – that contains fast, high-fidelity gates that can be executed simultaneously across a two-dimension qubit array whose qubits are connected by adjustable ("configurable"/"programmable") couplers (see Figure 1).

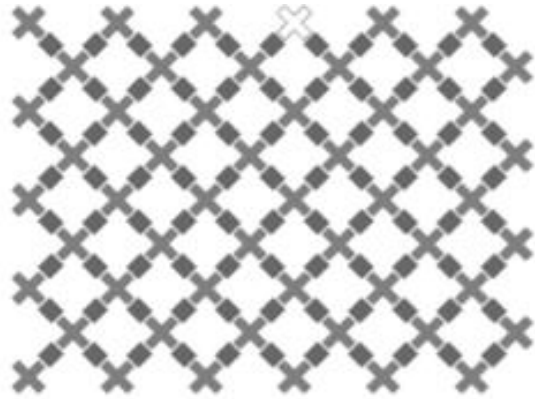

**Figure 1. Connection-scheme of a 53 (working)-qubit Sycamore processor (adapted from Arute et al. 20190a, 505). Solid crosses denote qubits that work; outlined crosses denote qubits that don't work. "Squares" represent "adjustable couplers".**

We will call a processor that has the kind of connection-scheme shown in Figure 1 a "Sycamore-architecture processor".

The Google Quantum Team evaluated the correctness of Sycamore's computation in two general ways: (a) by comparison with the output of a supercomputer running simulations of random quantum circuits, and (b) by component-level calibration of the Sycamore system, extrapolated to all configurations of interest. Some details on each of these approaches follows.

(a) The bitstring output of Sycamore's simulation of sampling of pseudo-random entangled quantum circuits was compared with the output of a classical supercomputer (the JUQCS-E and JUQCS-A simulators (De Raedt et al. 2018) on supercomputers at the Jülich Supercomputing Center) simulating the sampling of entangled pseudo-random quantum circuits. For simulating circuits of 43 qubits or fewer, the supercomputer used in the experiments employed a Schrödinger algorithm (Arute et al. 2019b, 49-50) to compute the full quantum state.

For pseudo-random quantum circuits containing more than 43 qubits, Arute et al. 2019a report there was not enough random access memory in the supercomputer used in the experiments to store the full quantum state, so the full quantum state had to be calculated by other means. On the supercomputer used, in order to simulate the state of pseudo-random quantum circuits containing 44-53 qubits, Arute et al. 2019a employed a hybrid Schrödinger-Feynman algorithm (Markov et al. 2018). The latter method breaks a circuit into two (or more) subcircuits, simulates each subcircuit with a Schrödinger method, then "reconnects" the subcircuits using an approach in some ways analogous to the Feynman path integral (Grosche 1993).

The outputs obtained from the Sycamore processor and the supercomputer simulations agree well for the specific set of simulated quantum circuits analyzed by Arute et al. 2019a (Figure 4a).

(b) To further assess the results of comparison of the outputs of the Sycamore processor and the corresponding supercomputer simulations ((a), above) the Google Quantum Team calibrated the "fidelity" of the Sycamore processor for a subset of possible circuit configurations of 53 or fewer qubits. The Google Quantum Team showed that a "fidelity" model (Arute et al. 2019b, Section

VIII) defined in terms of component-level error rates in a Sycamore-architecture processor containing Q qubits

$$F = \prod_{g \in G_1} (1 - e_g) \prod_{g \in G_2} (1 - e_g) \prod_{q \in Q} (1 - e_q),$$

(Eq. 1)

where there are $G_1$ gates of "Type 1" and $G_2$ gates of "Type 2", and
    $e_g$ is the Pauli error of gate $g$
    $e_q$ is the state preparation and measurement error of qubit $q$
    $F$ is the fidelity of the processor configuration of interest

produces predictions that agree well with the fidelity Sycamore-architecture systems used in the Google Quantum Team's experiments. The predictions of Eq. 1 thus at least partially corroborate the results of (a), for the cases tested by the Google Quantum Team.

If extrapolated, Eq. 1 might seem to be a fidelity/error-distribution model for Sycamore-architecture systems containing more than 53 qubits, or for circuit configurations of 53 qubits or less that were not simulated. To this, and the more general question of whether GQS can be extrapolated to cases not yet observed, we now turn.[3]

### 3.0 What do the Google Quantum Team's experiments show about quantum supremacy?

The Google Quantum Team's methods and results make a compelling case that GQS summarizes the experiments the Team performed. Can GQS be extrapolated to cases *not* tested to date? Answering that question requires answering several others. Among these are

1. Could an extrapolation of Eq. 1 for Sycamore-architecture systems containing more than 53 qubits be grounded in statistical inference theory, based on the results reported in Arute et al. 2019?

2. Does any method for comparing the performance of a quantum computer to the performance of a classical computer generalize beyond cases tested (including, but not limited to, those tested by the Google Quantum Team)?

### 3.1 Could an extrapolation of Eq. 1 for Sycamore-architecture systems containing more than 53 qubits be grounded in statistical inference theory, based on the results reported in Arute et al. 2019?

---

[3] Arute et al. 2019 do not explicitly claim or imply that Eq. 1 or GQS can be extrapolated for configurations other than those tested by them.

Let's assume that the Google Team's experiments show that Eq. 1 produces, for Sycamore-architecture systems containing no more than 53 qubits, fidelity predictions that agree well enough with those of the simulators running on classical computers.

The mathematical form of Eq. 1 is identical to the mathematical form of the fidelity/error distribution of a collection of components whose error distributions are independent and identically distributed (IID; (Hogg, McKean, and Craig 2005, 121)). IID models are commonly used to characterize system-level failure distributions in such as computer components, lighting equipment, fasteners, and many other mass-produced hardware items (O'Connor 2002). One might be tempted to infer that Eq. 1 can therefore be unproblematically extrapolated to describe the distribution of errors in Sycamore systems containing more than 53 qubits. But is this reasoning robust?

There is no question that the results of the Google Quantum Team's experiments are *consistent with* the behavior of a Sycamore-architecture system containing fewer than 54 qubits whose component-level error distributions are characterized by IID. Eq. 1, however, *does not imply* (in the sense of Russell and Whitehead 1910, 98) the component-level error distributions are IID. Why? Consider a set of component-level non-IID that in aggregate produce exactly the predictions of Eq. 1 for systems of 53 or fewer qubits. Let's call such cases Kalai configurations.[4]

The existence of Kalai configurations tells us that we have to be clear what role Eq. 1 is intended to play if it is extrapolated beyond the experimental evidence. Any such extrapolation would have to treat Eq. 1 as a distribution function in a *statistical inference* (Hogg, McKean, and Craig 2005, Chaps. 5-12).[5] Statistical inference requires that its domain of application be characterizable in terms of random variables. More specifically,

> (AP) Random variables are defined in terms of random experiments. In order to define a random experiment, E, we must know, *independent of experience* (i.e., *a priori*), all possible outcomes of that experiment (Hogg, McKean, and Craig 2005, Section 1.1 and Df. 1.5.1).

As used by Arute et al. 2019, Eq. 1 concerns only physics-/hardware-related errors. There is another potential source of error in computing systems that is not as such derivable from the physics or hardware of the system as such. In the experiments of Arute et al. 2019 in particular, the process of configuring a Sycamore architecture, a prescription for "adjusting the couplers" involves manipulating the apparatus in ways that are formally equivalent to "writing a program" or "developing software" for the processor. For economy of expression, we will call such a prescription *software*. Software in this sense can include, among other things, conditional prescriptions. For example, suppose that the couplers in a Sycamore-architecture system are named A, B, C… and suppose also that we prescribe that the coupler between A and B must be "adjusted" *if* B is coupled to C. This kind of conditional prescription would instantiate a binary branch in the set-up/software of the system.

Let S be software in the sense of the previous paragraph. Suppose S contains, on average, one binary branch per 10 instructions. Branching induces an execution-path network (equivalently, an execution state-space (Valmari 1998)) structure on S. In order to know – as required by AP -- *all* the possible outcomes of a random experiment E performed on S, we must know what S will do when each of these paths is executed. Now in general, the number of such paths in a system of M instructions, with, on average, a binary branch instruction every N instructions, is $2^{M/N}$, where N < M. In general, determining S's behavior on all paths by *testing* would take ~$10^{13}$ lifetimes of the universe even if M is as small as 1000 and N = 10 (see Symons and Horner 2014; 2017; 2019 for further detail).

---

[4] This is a generalization of a suggestion made by Gil Kalai (See Israeli Institute for Advanced Science 2019).

[5] For further detail, see Symons and Horner 2017.

The testing-based "state-explosion" (Valmari 1998) problem sketched in the previous paragraph could be overcome if we could determine, without testing (i.e., a priori), the behavior of S on each path. In particular, if we had a method for automatically generating provably correct software from provably correct models/specifications, and if using this method were less onerous than exhaustively testing all paths in the resulting software, it would seem that we could dodge the state-explosion problem faced by characterization-through-testing.

There is such a method, generically called *model checking*, and it has contributed to some impressive practical results (Clarke et al. 2018). Notwithstanding those achievements, however, almost if not all software involving model checking to date has been produced by using software (development) systems/environments that themselves were not produced through model checking. Those software development environments typically include editors, compilers, linkers, operating systems, etc., and each of these contains thousands to millions of instructions. To the extent that these environments were not developed through model checking, we cannot infer that the software produced by using those environments is fully characterized (see Symons and Horner 2017).[6]

We conclude, therefore, that Requirement AP cannot be satisfied by purely empirical methods. This implies, in particular, that Eq. 1 and GQS cannot, on the basis of experiments like those of Arute et al. 2019 per se, be extrapolated by statistical inference to cases that have not been observed.

### 3.2 Does any method for comparing the performance of a quantum computer to the performance of a classical computer generalize beyond cases tested?

In order to answer this question, we need to be clear about what it means to compare the performance of two physical computing systems. We posit the following condition of adequacy for any such comparison:

> (CCA) Let A and B be physical computing systems. Let C(A) be a computation performed on A and C(B) be a computation performed on B. Let T(A) be the physical theory required to completely describe the trajectory of A while performing C(A). Let T(B) be the physical theory required to completely describe the trajectory of B while performing C(B). Then C(A) and C(B) can be compared only T(A) and T(B) are consistent with each other.

---

[6] The configuration-prescriptions used in the experiments of Arute et al. 2019 in particular were not produced in a software development environment all of which was produced by model checking.

The comparison of the performances of quantum and classical computers is not like comparing the performance of two classical computers that have the same architecture.[7] Why? By CQD, quantum computation requires that states can be superposed; classical computation, by definition, denies that state(s) can be superposed. Thus, by (CCA), classical and quantum computations are simply not comparable.

Given the Copenhagen Interpretation (CI) of quantum mechanics (see, for example, Messiah 1958, Vol. I, Chap. 4, Sections 16 and 17), furthermore, even determining the state of a quantum computer is inherently different from determining state the state of a classical computer: in the CI, quantum-state-measurement is relativized to specific measurement configurations, but classical-state-measurement is not. Thus, given the CI, by (CCA) classical and quantum computations are not comparable.

Given these arguments, what can the claim that the performances of a classical, and a quantum computer, at least in one case, are comparable (as, for example, Arute et al. 2019 do) mean? Just this: any such a claim is only comparing the speeds at which a given classical and a given quantum computer produce given classically described outputs, given classically described inputs. That kind of claim, however, treats a computer as a (classical) "black box". By definition, black-box comparisons compare only outputs, for given inputs (Patton 2005). A black-box comparison is defined only for particular pairs of outputs because, by definition, there *can no theory that allows us to extrapolate* from the particular observations we have made to anything else.[8]

## 4.0 Conclusions

Arute et al. 2019 provide strong empirical evidence of GQS for the test cases considered. Given CQD in Section 1.0, Requirement AP in Section 3.1, however, quantum/classical computing comparisons can show at most *only* that GQS holds for the set of specific quantum/classical-computer pairs that have already been observed. Given the problems of comparing classical and quantum computers discussed in Section 3.2, furthermore, it is not possible to compare classical and quantum computing except in a "black-box" sense, and such comparisons cannot be extrapolated beyond the results of specific experiments. These results show that generalizing the claim of quantum supremacy beyond what has been observed in particular cases will continue to be a difficult problem.

## 5.0 Acknowledgements

---

[7] Even defining what "same architecture" means in the case of classical computers is complicated (see Hennessey and Patterson 2007; Jagode et al. 2019).

[8] This problem is *not* the same as the well-known problem of induction (see, for example, Hume 1739, Book I, Part III; Salmon 1966, Section I), i.e., the problem of inferring events we haven't observed from those we have. The quantum/classical computer comparison conundrum trades only the fact that the concepts of "state" in quantum and classical contexts are incompatible.

We would like to thank the referees for this paper for their helpful criticism. Thanks also to Francisco Pipa for his careful reading and comments. JFS was supported in part by NSA Science of Security initiative contract #H98230-18-D-0009. JKH was supported in part by a Ballantine Foundation grant.